\documentclass[twocolumn]{emulateapj}

\usepackage{graphicx}
 \usepackage{amsmath}
\usepackage{txfonts}
\usepackage{longtable}
\usepackage{lscape}
\usepackage{natbib}
\usepackage{soul}

\makeatletter
\AtBeginDocument{\let\LS@rot\@undefined}
\makeatother

\usepackage[usenames,dvipsnames,svgnames,table]{xcolor}

\shorttitle{Velocity difference of ions and neutrals in solar prominences}

\shortauthors{Wiehr, Stellmacher, Balthasar \& Bianda}

\begin{document}

\title{Velocity difference of ions and neutrals in solar prominences}

\author{E. Wiehr\altaffilmark{1,}, 
  G. Stellmacher\altaffilmark{2},
  H. Balthasar\altaffilmark{3}, and 
  M. Bianda\altaffilmark{4}} 

\altaffiltext{1}{Institut f\"ur Astrophysik, D-37077 G\"ottingen, Germany}
\altaffiltext{2}{Institut d'Astrophysique, F-75014 Paris, France}
\altaffiltext{3}{Leibniz-Institut f\"ur Astrophysik, D-14482 Potsdam, Germany}
\altaffiltext{4}{Istituto Ricerche Solari Locarno, Universit\`a della Svizzera italiana, 
CH-6605 Locarno, Switzerland}

\email{ewiehr@gwdg.de, hbalthasar@aip.de, mbianda@irsol.ch}

\begin{abstract}
   
  Marked velocity excesses of ions relative to neutrals are obtained from two
  time series of the neighboring emission lines He\,\textsc{i}\,5015\,\AA{}
  and Fe\,\textsc{ii}\,5018\,\AA{} in a quiescent prominence. Their Doppler
  shifts show time variations of quasi-periodic character where the ions are
  faster than the neutrals
  $1.0\le V_{macro}($Fe\,\textsc{ii}$)/V_{macro}($He\,\textsc{i}$)\le1.35$
  in series-A and, respectively, $\le1.25$ in series-B. This 'ratio excess'
  confirms our earlier findings of a 1.22 ion velocity excess, but the present
  study shows a restriction in space and time of typically 5\,Mm and 5\,min.

The ratio excess is superposed by a time and velocity independent 'difference excess'
$-0.3\le V_{macro}($Fe\,\textsc{ii}$)-V_{macro}($He\,\textsc{i}$)\le+0.7$\,km\,s$^{-1}$
in series-A (also indicated in series-B). The high repetition rate of 3.9\,sec enables
the detection of high frequency oscillations with several damped 22\,sec periods in
series-A. These show a ratio excess with a maximum of 1.7. We confirm the
absence of a significant phase delay of He neutrals with respect to the Fe ions.
\end{abstract}


\section{Introduction}
\label{Sect:Introd}

Charged particles may be faster than neutrals in a partially ionized and weakly
collisional magnetized plasma (cf. review by Ballester et al., 2018). This ion
drift can be described by a multi-fluid model, where the different species interact
by weak coupling processes (Gilbert et al. 2002). For a detailed study of this effect,
quiescent solar prominences offer a particularly good example, since they allow higher
resolution in space and time than stellar objects. The observed emission of lines
from neutrals, as He\,\textsc{i}, shows that the prominence plasma is not fully ionized.

A velocity excess of ions over neutrals in prominences was observed and discussed in
recent papers. Whereas Khomenko et al. (2016) found such an excess in restricted
prominence areas with high velocities of short-lived transients, Wiehr et al. (2019)
found systematically larger ion shifts through an evolutionary stable prominence.
Besides larger Doppler shifts, prominence emissions also show broader line widths
for ions than for neutrals (Ramelli et al., 2012, Fig.5; Stellmacher and Wiehr,
2015, Fig.1). This suggests an excess of non-thermal broadening by higher ion
velocities on a very small scale.

Here, we study the excess of macro velocity of ions over neutrals in time and
space from high resolution time series. We extend and improve former observations
of the alternately measured Na$\,\textsc{i}\,5896$\,\AA{} (D$_1$) and
Sr$\,\textsc{ii}\,4078$\,\AA{} lines by simultaneous observations of the neighboring
lines He$\,\textsc{i}\,5015.7$\,\AA{} (singlet) and  Fe$\,\textsc{ii}\,5018.4$\,\AA{}.
Their small wavelength distance of 2.7\,\AA{} avoids influences by the refraction in
Earth's atmosphere, which are a relevant problem for time-series observations, since
the direction of refraction rotates with respect to the solar disk coordinates (cf.
Wiehr et al., 2019).

Parasitic light, superposing the prominence emission with an absorption spectrum,
shows different rotational Doppler shift than the prominence emission lines (cf.
Wiehr et al., 2019). This may introduce problems if a reference spectrum for the
parasitic light cannot be taken in the immediate neighborhood, as for laterally
extended prominences. 

\section{Observations}

We observed a laterally small prominence at the east limb, $5^\circ$ North on June 28,
2019, from 7:45\,UT through 8:25\,UT (Fig.1) with the 45 cm aperture Gregory-Coud\'e telescope
and its Czerny Turner spectrograph (f=10\,m) of the swiss observatory Istituto Ricerche 
Solari Locarno. A fixed slit of correspondingly 1.5\,arcsec width (1000\,km on the Sun)
was oriented perpendicular to the solar limb. 

\begin{figure}[!h]
\centering
\includegraphics[width = 0.48\textwidth]{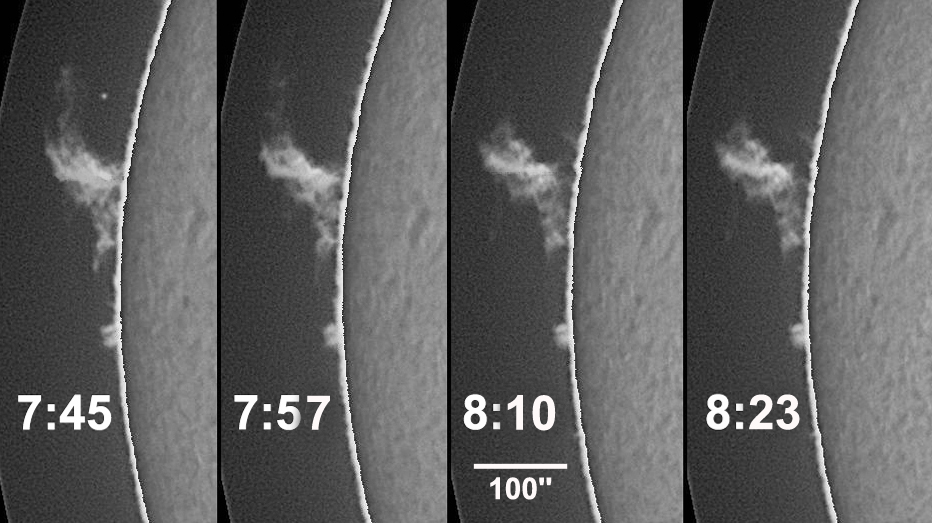}
\caption{H$\alpha$ image sequence of the prominence at the east limb, $5^\circ$ north
  from June 28, 2019, (Learmouth observatory); north direction is upwards; the 100''
  bar corresponds to 73.4\,Mm.}
\label{Fig1} 
\end{figure} 

We took time series of the emission lines He$\,\textsc{i}\,5015.7$\,\AA{} and
Fe$\,\textsc{ii}\,5018.4$\,\AA{}. Their brightness allowed a repetition rate of 3.9 sec,
which is 11 times faster than that of Wiehr at al. (2019), the spatial resolution being
almost the same. Series-A with 105 spectra (7:45-7:52\,UT) shows smooth emissions of He
and Fe$\,\textsc{ii}$ along the slit (upper part of Fig.\,2). Series-B with 460 spectra
(7:55-8:25\,UT) shows brighter emissions with marked Doppler shifts and evolutionary
time variations (well visible in the lower part of Fig.\,2). Sufficiently bright emissions
of the faint lines occur in height levels of 31\,Mm - 50\,Mm above the solar limb, the
31\,Mm boundary marking the lower edge of the prominence main body seen in H$_{\alpha}$
(cf. Fig.\,1). Beyond spectrum 320 of series-B (7:57\,UT) this edge started to uplift with
$\approx9$\,km\,s$^{-1}$ (dashes in Fig.2), and the spectra show increasingly fragmented
emissions with line satellites, announcing the sudden disappearance at 9:45\,UT.

\begin{figure}[!h]
\centering
\includegraphics[width = 0.465\textwidth]{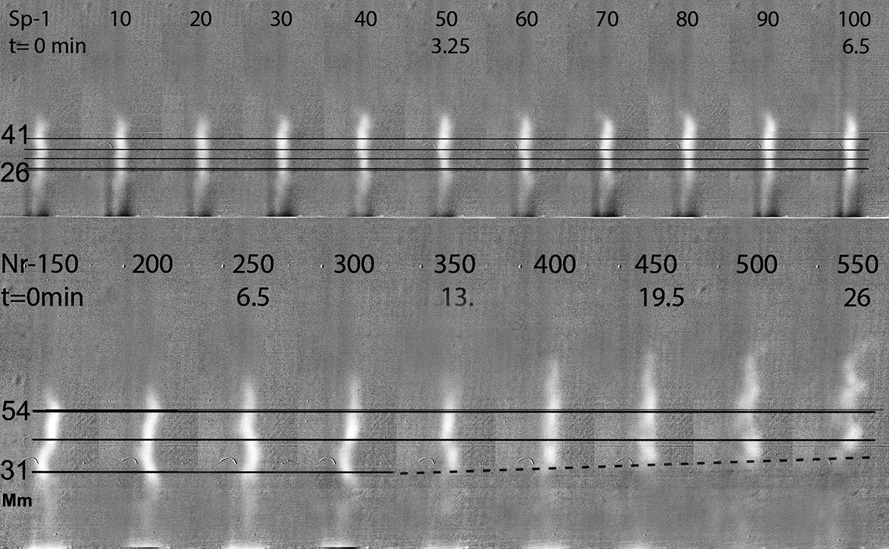}
\caption{Time sequence of the Fe$\,\textsc{ii}\,5018.4$\,\AA{} emission spectra in
  prominence E5N for series-A (upper) and B (lower image part). Scan rows at different
  height above the limb are marked by full lines; for series-B the lower emission boundary
  lifts up with $\approx9$\,km\,s$^{-1}$ (dashed line) after 13\,min of observation
  (spectra with no.$\le320$); the spatially fragmented emissions in the last two spectra
  (500 and 550; recorded at 8:18 and 8:21) announce the sudden disappearance at 9:45\,UT.}
\label{Fig2}
\end{figure}

\section{Data reduction}

In order to remove parasitic light superposing the emission lines we use spectra of
the 'aureola' taken immediately before and after the prominence time series at a slit
position in the immediate prominence neighborhood. (For details of the reduction procedure
see Ramelli et al., 2012). We average 9 CCD rows, each 257\,km wide (0.35\,arcsec),
to adapt the effective spatial resolution to the slit width. 

As wavelength reference for Doppler shifts we determine the centers of the absorption
lines Ti\,I\,5016.1, Fe\,I\,5016.9 and Ni\,I\,5017.5, before removing the superposed
parasitic light in each individual spectrum, fitting polynomials of second degree. This
wavelength reference from the aureole is independent of complex photospheric velocity
fields (e.g. oscillations and granular blue shift), which interfere if disk center
spectra are used for a wavelength calibration.

We correct the velocities of each individual spectrum for the shift of the three
absorption lines relative to their position in the aureola spectra. The resulting
macro-velocities are thus free from spectrograph drifts and from slow terms of
spectrograph seeing. The two slit positions chosen for the emission spectra and for
the aureola spectra aside the prominence, have slightly different inclinations to
the solar limb. Hence, their absorption lines show slightly different variation of
rotational Doppler shifts along the slit. However, this effect is negligible since
the prominence is small and the two slit positions are thus rather close. 

The wavelength reference is taken at heights between 40'' and 75'' above the limb,
where the aureole spectrum is 1.4\,km\,s$^{-1}$ less blue-shifted than the rotational
shift of the east limb (for details see Wiehr et al., 2019). We correspondingly correct
the deduced Doppler shifts, thus referring to a co-rotating reference system relative to
the photosphere below the prominence. 

\begin{figure}[!t]
\centering
\includegraphics[width = 0.495\textwidth,]{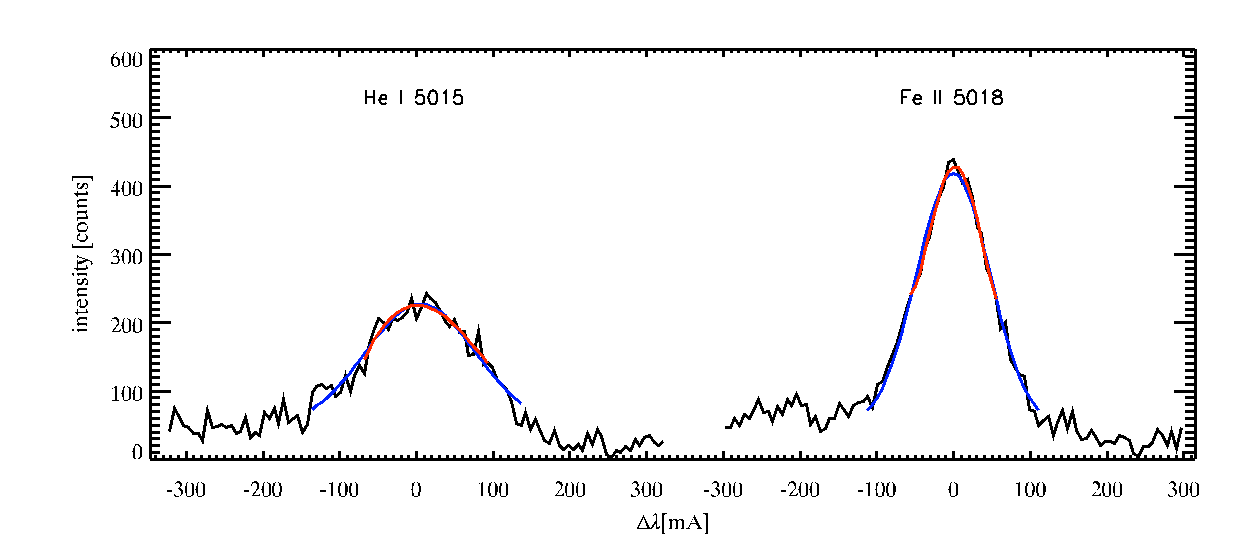}
\caption{Example of the finally obtained emission profiles of
  He$\,\textsc{i}\,5015.7$\,\AA{} and Fe$\,\textsc{ii}\,5018.4$\,\AA{}. Blue curves
  give the Gaussian, red curves the polynomial fit.}
\label{Fig3}
\end{figure}

For the determination of macro-shifts and full widths at half maximum we fit single
Gaussians to the upper part of the emission profile. Alternately, we fit polynomials
of fourth degree, which better represent asymmetric line profiles. The macro velocities
obtained by both methods differ by less than 1\%, indicating that the observed profiles
of both emission lines are indeed largely symmetric. We therefore use the Gaussian fits,
which give an estimated accuracy of $\approx20$\,m/s for the macro velocities

Fig 3 shows an example of finally obtained emission profiles. That of
Fe$\,\textsc{ii}$ is narrower due to the 14 times higher mass of Fe. The noise is
low enough to ensure reasonable fits.

\section{Results}

\subsection{Time series-B}

Doppler-time sequences of Fe$\,\textsc{ii}\,5018.4$ (full lines) and He$\,\textsc{i}\,5015.7$
(dashed lines) are shown in Fig.\,4 for a 4-spectra mean (15.6\,s resolution) of the first
17\,min of series-B. The macro shifts give quasi-periodic velocity variations of several km\,s$^{-1}$
with almost equal amplitudes for both lines. They are superposed by a velocity difference
of Fe$\,\textsc{ii}$ relative to He$\,\textsc{i}$. This 'difference excess' is  nearly
constant with time (i.e. spectrum number), independent on the velocity itself and
diminishes at larger heights.

\begin{figure}[!h]
\centering
\includegraphics[width = 0.495\textwidth]{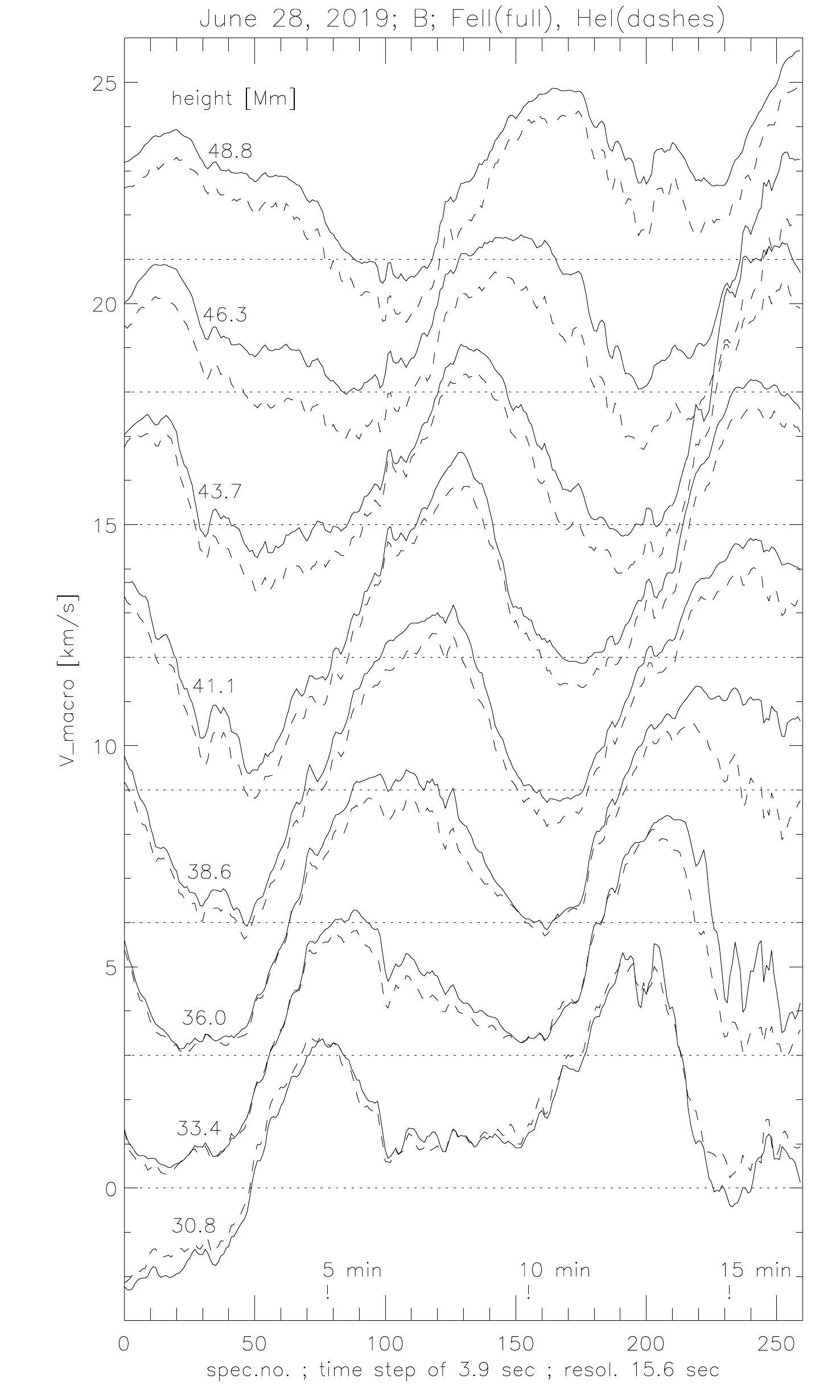}
\caption{Macro velocities of Fe$\,\textsc{ii}\,5018.4$\,\AA{} (full) and
  He$\,\textsc{i}\,5015.7$\,\AA{} (dashed) relative to the photosphere below the prominence
  for series-B; each scan averages over 9 CCD rows (i.e. 2.3\,Mm resolution); time scale
  integrated over 4 spectra, (15.6\,s time resolution); scans shifted to each other by
  3\,km\,s$^{-1}$ (dotted lines); height levels are indicated.}
\label{Fig4}
\end{figure}

\begin{figure}[!h]
\centering
\includegraphics[width = 0.495\textwidth]{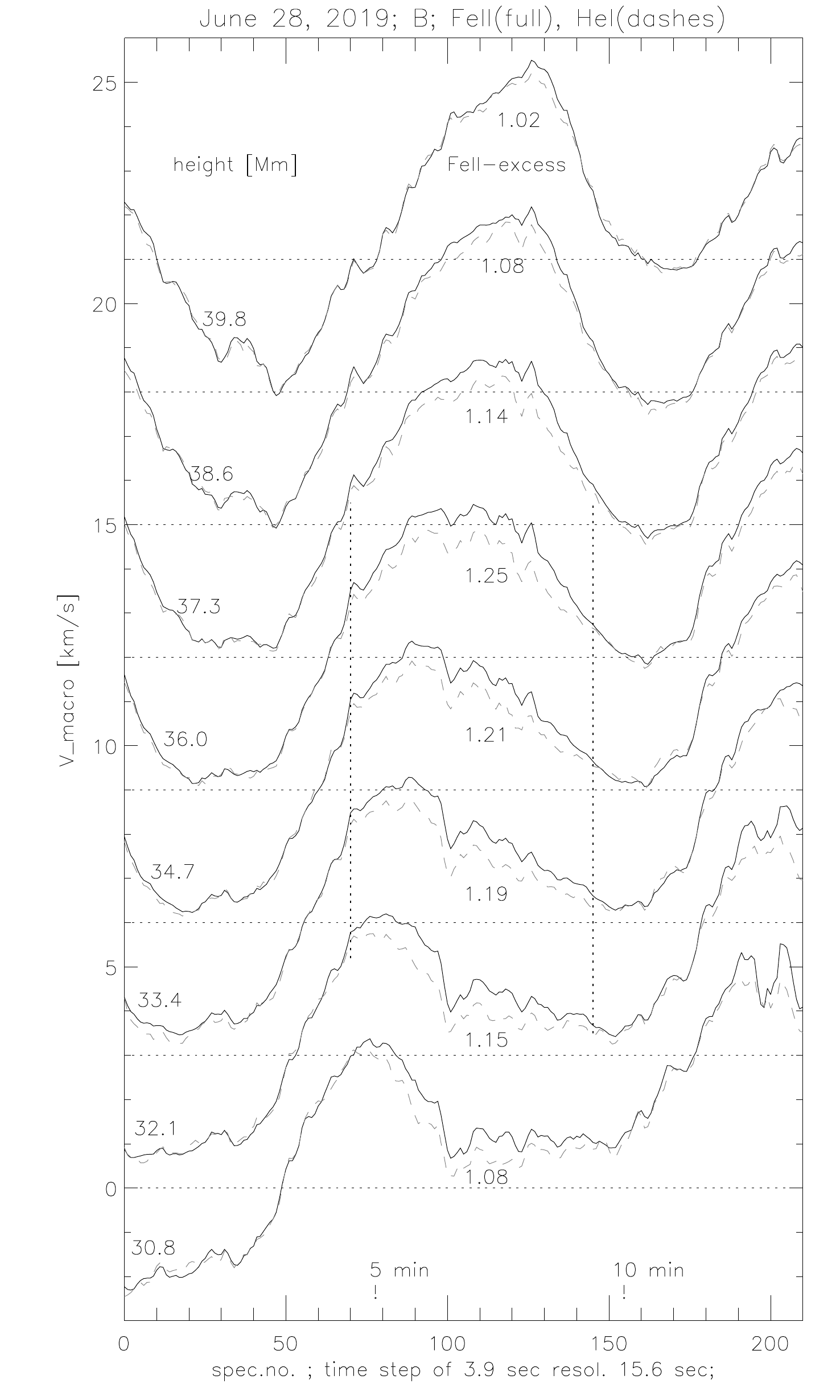}
\caption {Same as Fig.4, but the He$\,\textsc{i}$ velocity (dashes) shifted to fit that
  of Fe$\,\textsc{ii}$ over a smaller height range in the first 70 spectra of series-B;
  the scan rows are vertically displaced by +3\,km\,s$^{-1}$; dotted vertical lines mark
  the limits of time and space regimes with excess velocity ratio $>1.1$; height levels
  are indicated.}
\label{Fig5}
\end{figure}

\begin{figure}[!h]
\centering
\includegraphics[width = 0.495\textwidth]{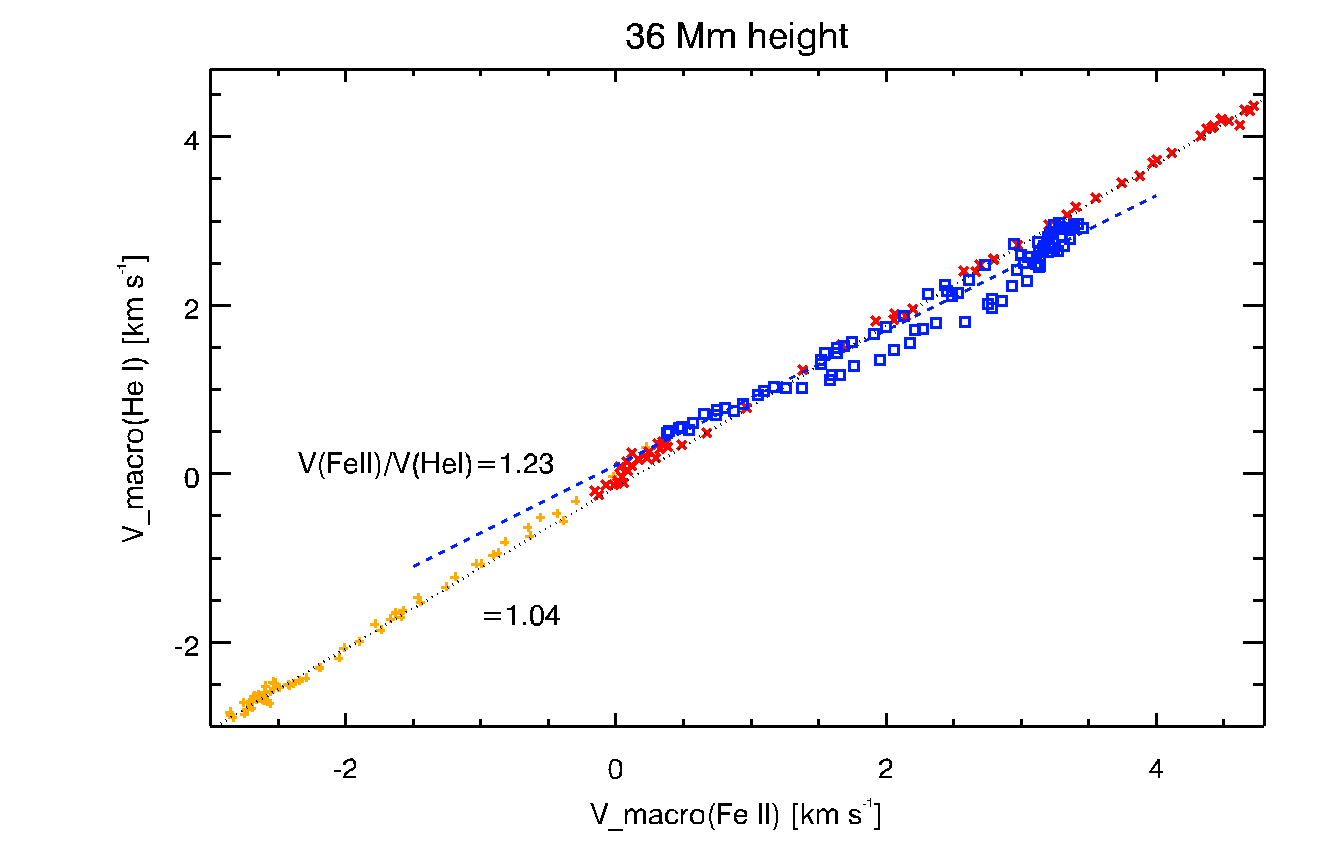} 
\caption {Scatterplot of He$\,\textsc{i}\,5015.7$ and Fe$\,\textsc{ii}\,5018.4$
  velocities of series-B at the height h$=36\pm1$\,Mm with maximum ion velocity
  excess; spectra-0-69 (yellow\,+) and 151-210 (red\,X) give no excess;
  spectra-70-150 (blue\,$\square$) give an excess of 1.23.}
\label{Fig6}
\end{figure}

If we remove the time-constant ion drift, most of the differences between the
Fe$\,\textsc{ii}$ and the He$\,\textsc{i}$ velocities disappear (Fig.\,5), reflecting  
the largely equal amplitudes. However in the time interval of minute 4.5 through 
9.5 of series-B a superposed excess of Fe$\,\textsc{ii}$ velocities becomes visible
at heights $\le40$\,Mm above the limb. In contrast to the 'difference excess', 
this excess is related to the velocity itself and thus a 'ratio excess'. Scatterplots 
(e.g. Fig.\,6) show that it smoothly varies with height between  $32 < h < 37$\,Mm 
and disappears for $h > 40$\,Mm. Its maximum amounts to 
V$_{macro}$(Fe$\,\textsc{ii}$)/V$_{macro}$(He$\,\textsc{i}$)=1.25, in agreement 
with Wiehr et al. (2019).

In Fig.\,7 we show the time variation of the difference of Fe$\,\textsc{ii}$ and 
He$\,\textsc{i}$ velocities. [Note that the ordinate scale is about five times
expanded as compared to Figs.\,4 and 5.] The nearly constant values through the first
4.5\,min again reflect the largely equal velocity amplitudes of Fe$\,\textsc{ii}$ and
He$\,\textsc{i}$. We mark the shifts applied to transform Fig.\,4 into Fig.\,5 by
dashed lines and vertical arrows. These vary from $-0.35$\,km\,s$^{-1}$ ('blue excess')
at h=30.8\,Mm to $+0.6$\,km\,s$^{-1}$ ('red excess') at h=39.8\,Mm above the limb,
yielding an almost perfectly linear gradient of 1\,km\,s$^{-1}$ over 11.6\,Mm.

\begin{figure}[!h]
\centering
\includegraphics[width = 0.495\textwidth]{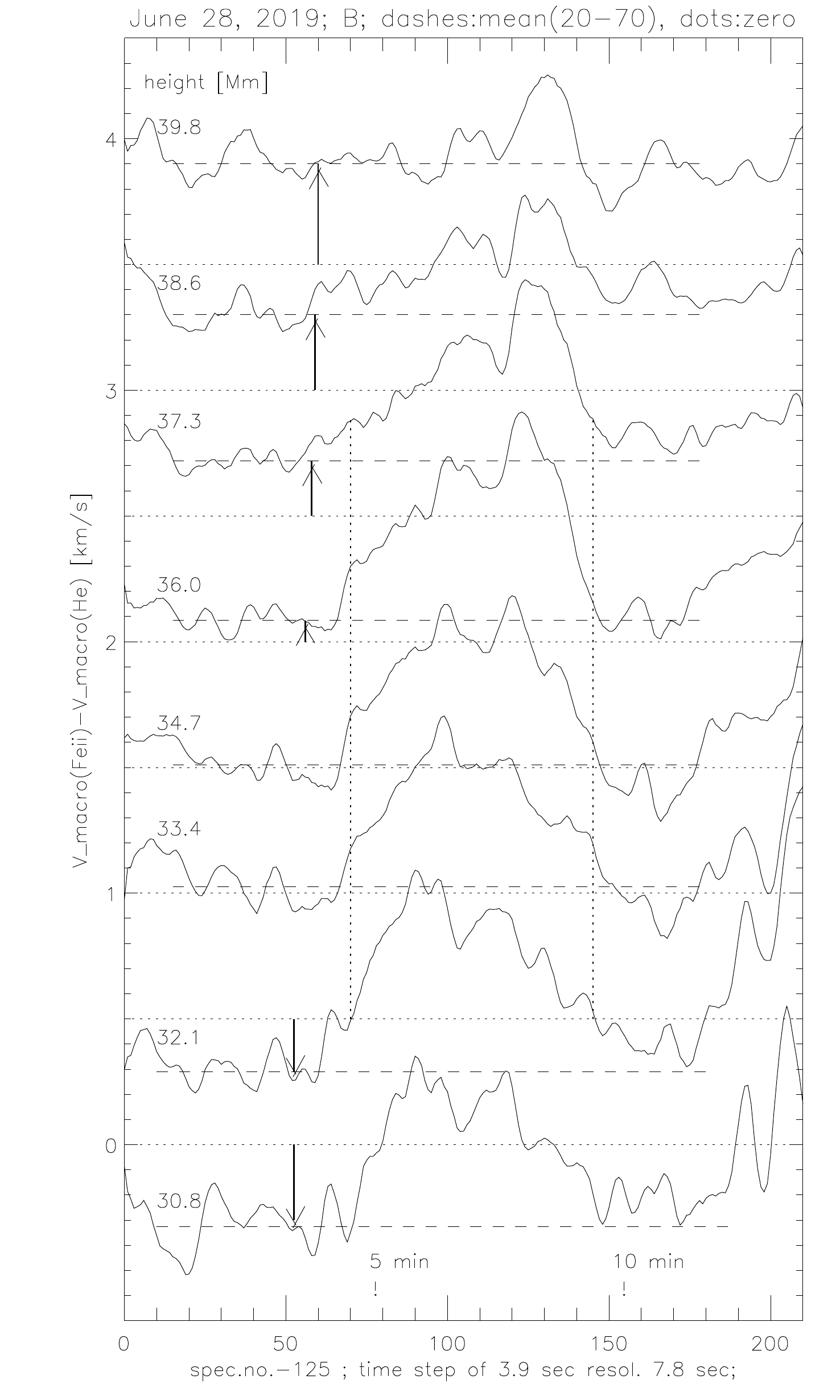}
\caption {Time variation of the velocity difference Fe$\,\textsc{ii}$ - He$\,\textsc{i}$,
  used to transform Fig.3 to Fig.4: horizontal dashed lines mark the offset taken from
  spectra-0 to 69: vertical arrows visualize the respective blue and red shifts.}
\label{Fig7}
\end{figure}

The onset of the Fe$\,\textsc{ii}$ ratio excess at spectrum-70 agrees with that in
Fig.\,5. We thus find two species of ion drifts: (i) a velocity difference excess
largely independent of the velocity itself (even through the marked velocity increase
between spectrum-40 and -70); and (ii) a ratio excess depending on velocity, and
restricted in space and time.

Beyond spectrum-210 ($>13.5$\,min of recording series-B) the macro velocity of the
He emission line stays increasingly behind that of the Fe ions (cf.\,Fig.\,4). The
corresponding spectra show marked fragmentation (cf. Fig.\,2) and motions along the
slit direction. Differences between Gaussian and poly fits indicate asymmetric
profiles caused by superposition of line satellites. This indicates the onset of
evolutionary velocity changes announcing the sudden disappearance. We thus restrict
our data analysis of series-B to the first 210 spectra.

\subsection{Time series-A}

The spectra of series-A are characterized by rather narrow and symmetric emission
lines (cf. Fig.\,2). Doppler-time sequences show quasi-periodic velocity
perturbations with amplitudes increasing with height from 0.35 to 1.45\,km\,s$^{-1}$
over the range 30 - 44\,Mm (Fig\,8). Above 40\,Mm we find a velocity excess 
$1.0\le V_{macro}($Fe\,\textsc{ii}$)/V_{macro}($He\,\textsc{i}$)\le1.35$. At lower
heights (h$\le39$\,Mm) no ratio excess is observed, however, a general shift
(`difference excess') is indicated in the lowest three scan rows at $h\le36$\,Mm
(cf.\,Fig\,8).

\begin{figure}[!h]
\centering
\includegraphics[width = 0.48\textwidth]{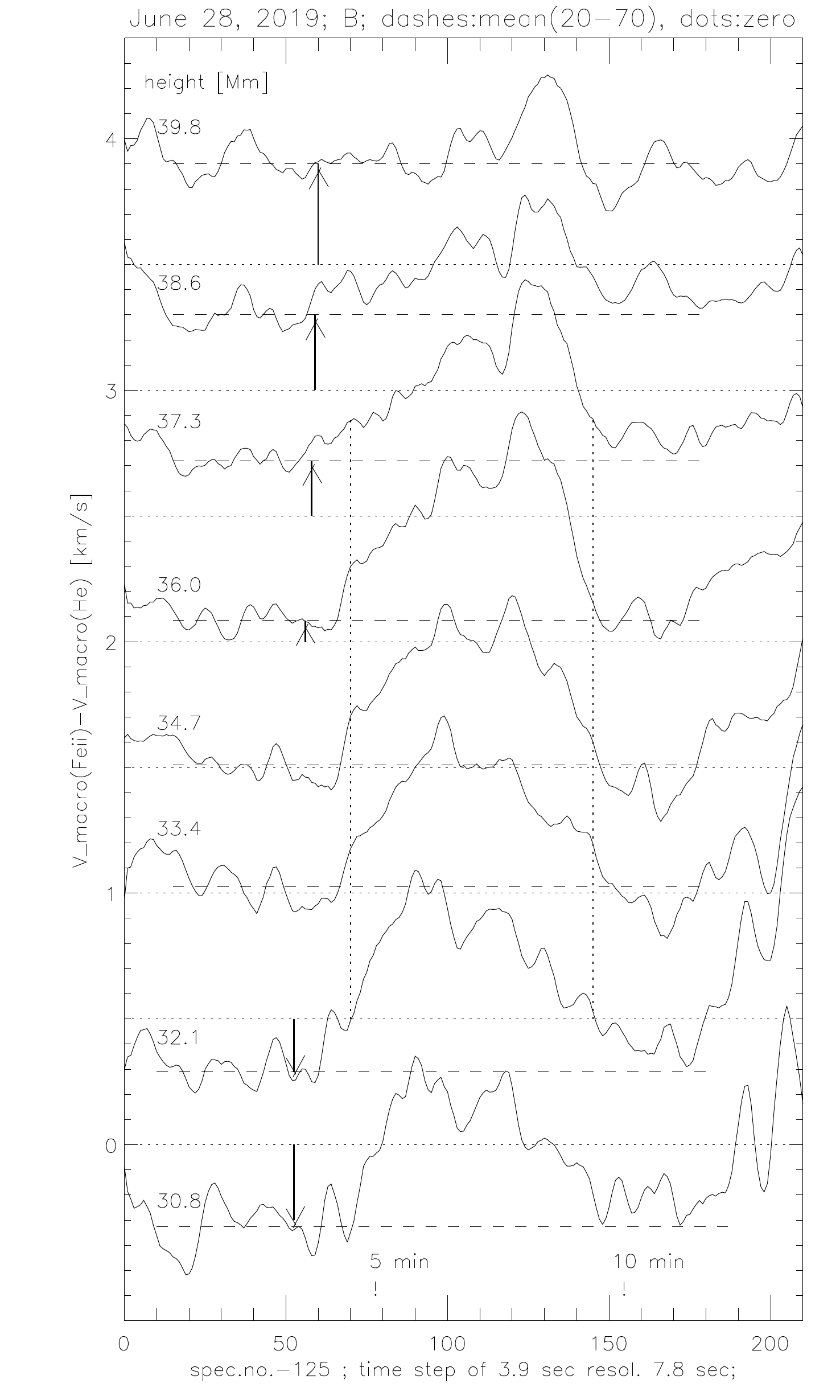}
\caption {Velocity variation of He$\,\textsc{i}$ (dashed) and Fe$\,\textsc{ii}$ (full line)
  in series-A; each scan averaged over 9 rows (2.3\,Mm resolution); time scale integrated
  over 2 spectra (7.8\,s resolution) in the height range 30.8-46.3\,Mm; the scans are
  shifted towards each other by 1.0\,km\,s$^{-1}$.}
\label{Fig8}
\end{figure}

\subsection{High frequency perturbations}

The high repetition rate of 3.9\,s enables the detection of short-period velocity
variations visible in the middle scan rows of spectra 40\,-\,90 in Fig.\,8. To make 
them visible, we suppress the long-term velocity perturbations, by subtracting the
10-spectra means from the 2-spectra means seen in Fig.\,8. The resulting
time series in Fig.\,9 shows at the height level 39.8\,Mm an almost perfect 
oscillation with decreasing amplitude through interval minute 2.6 to 5.5
(spectra 40\,-\,90).

\begin{figure}[!h]
\centering
\includegraphics[width = 0.495\textwidth]{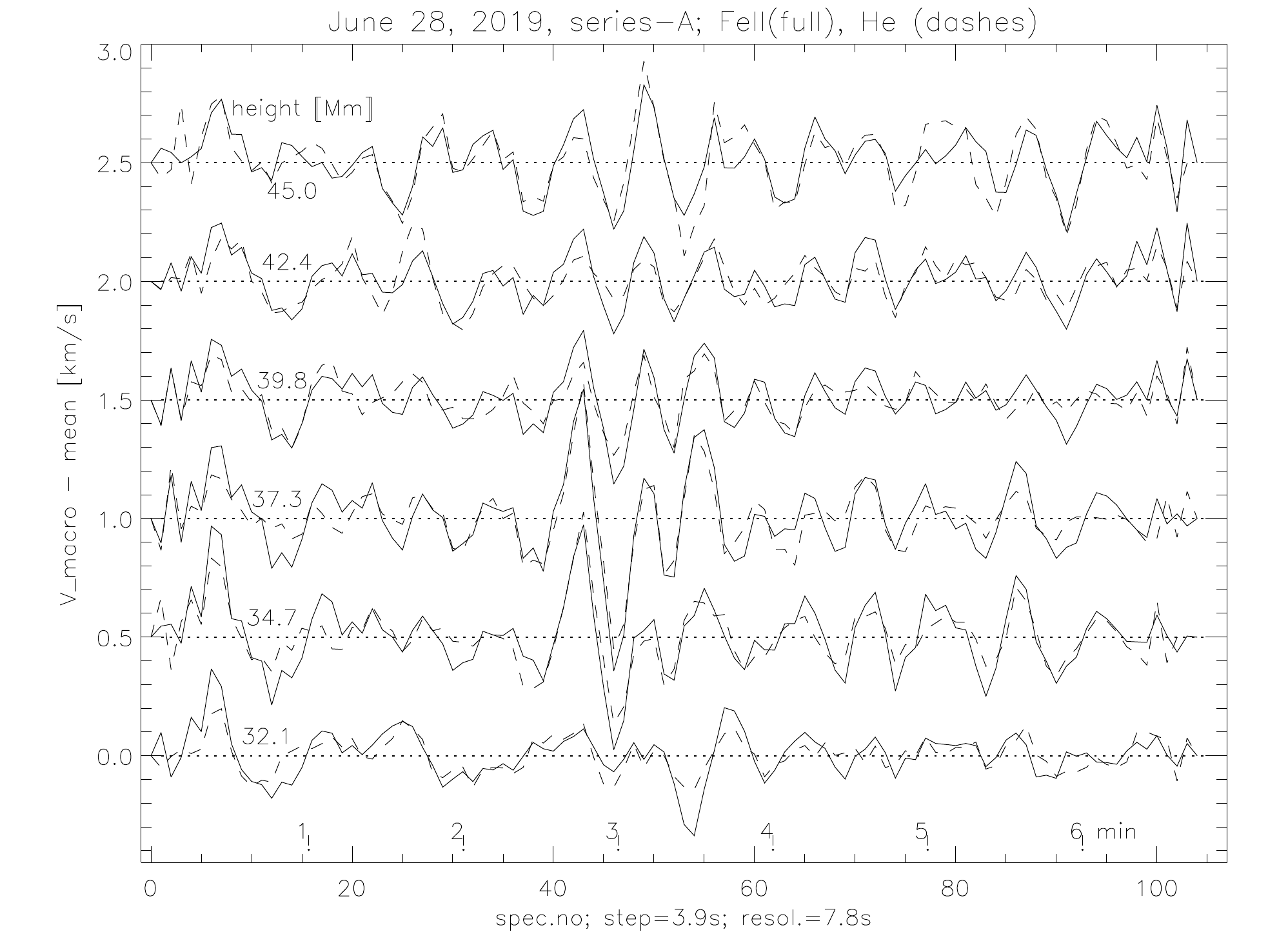}
\caption { Difference of macro shifts from 2 and 10 spectra means (7.8s and 39s time
  resolution) for the height range $30.8 \le h \le 43.7$\,Mm, covering the 22\,s
  oscillation region.}
\label{Fig9}
\end{figure}

The power spectrum Fig.\,10 of the corresponding height range $38<h<41$\,Mm over
the total series-A of 105 spectra shows a pronounced peak at 46\,mHz (22\,sec)
with a significance of 99.9\% for Fe$\,\textsc{ii}$ and 99.0\% for He$\,\textsc{i}$
(for details see Balthasar, 2003). Since the power maxima represent the square
of the velocity amplitudes, we obtain from the square roots of the power maxima of
the two emission lines their amplitude ratio. The velocity excess in the interval of
22\,sec oscillation amounts to 1.7. In the subinterval with pronounced oscillations
(spectra 40-90) scatterplots of the He$\,\textsc{i}$ and Fe$\,\textsc{ii}$ velocities
show that the velocity excess has a spatial variation almost parallel to that of the
large time-scale ratio of series-A (cf. section-4.2).

\begin{figure}[!h]
\centering
\includegraphics[width = 0.48\textwidth]{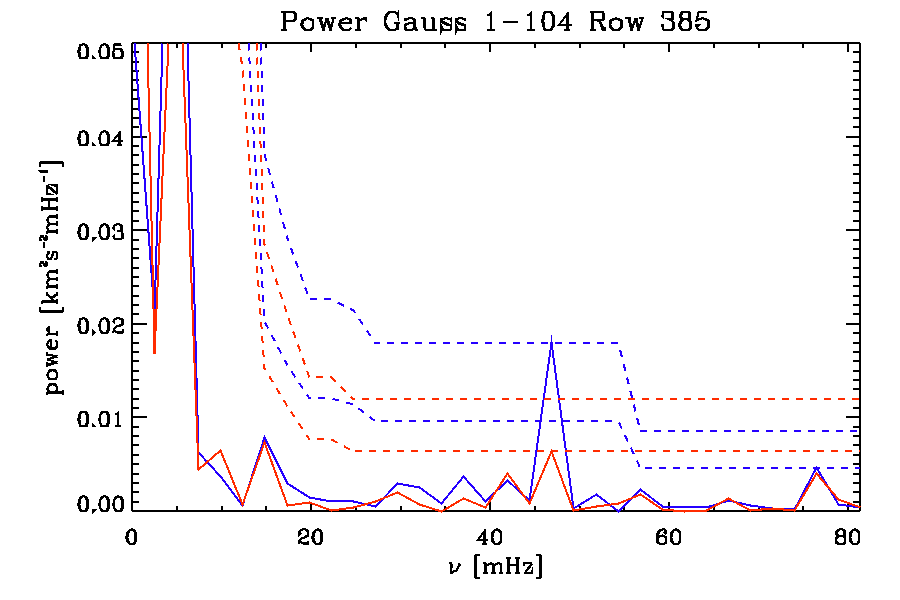}
\caption {Power spectrum of Fe$\,\textsc{ii}$ (blue) and He$\,\textsc{i}$ (red)
  at h=40\,Mm of series-A, showing a pronounced peak at 46\,mHz (22\,sec period);
  the 99\% and 99.9\% significance levels are indicated (dashed); the square root
  of the power maxima ratio of both emission lines gives a velocity excess of 1.7.} 
\label{Fig10}
\end{figure}

In series-B we also find high frequency variations with periods near 25\,s in
restricted intervals of space and time. These, however, show only few cycles and give
thus no significant peak in the power spectra. For the high frequency oscillations we
do not find a significant phase shift between Fe$\,\textsc{ii}$ and He$\,\textsc{i}$. 

\section{Discussion}

The observation of this prominence started $\approx$\,100\,min before its sudden 
disappearance. It shows two species of ion excess: (i) a velocity difference 
$V_{macro}($Fe\,\textsc{ii}$)-V_{macro}($He\,\textsc{i}$)$ and (ii) a velocity
ratio $V_{macro}($Fe\,\textsc{ii}$)/V_{macro}($He\,\textsc{i}$)$.
The marked 'difference excess' found in series-B increases linearly with height from 
-0.3\,km\,s$^{-1}$  (blue shift) at 30.9\,Mm height to +0.7\,km\,s$^{-1}$  (red shift) 
at 42.6\,Mm height (left panel of Fig.\,11). A similar effect was found in Wiehr
et al. (2019) where such difference excess increased between 3.3 and 39\,Mm from +0.4
to +2.1\,km\,s$^{-1}$. [Note that the velocities refer in both papers to the photosphere
below the prominence.] 

\begin{figure}[!h]
\centering
\includegraphics[width = 0.495\textwidth]{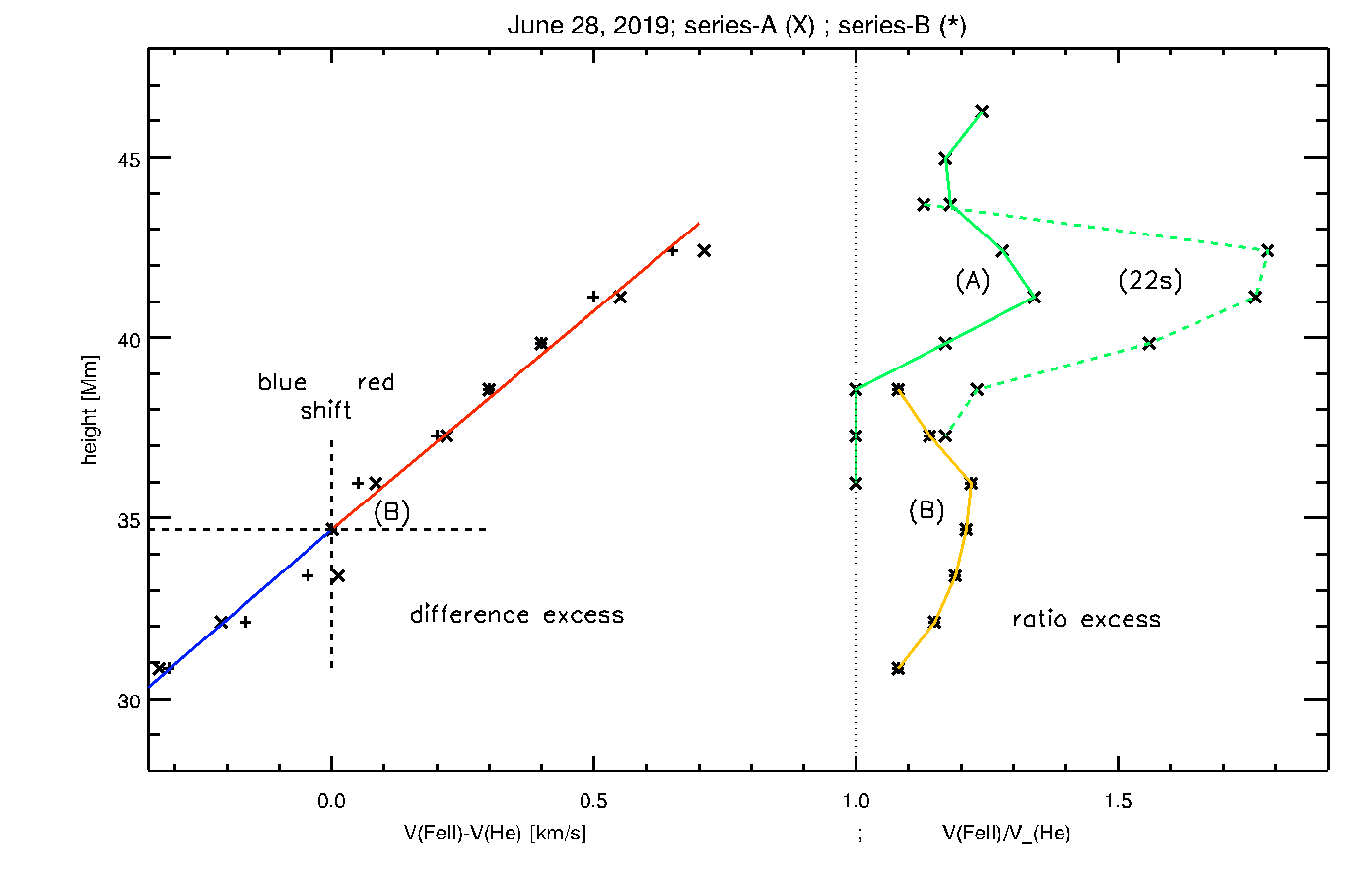}
\caption {Dependence of the difference excess (left) and the ratio excess (right
  side) on height above the solar limb for series-B (yellow line), series-A and its
  22\,s oscillation (green lines); difference excess determined visually (+) and by
  subtraction (X); its transition between blue- and red-shift occurs at 34.5\,Mm
  height above the limb (dash-line cross).} 
\label{Fig11}
\end{figure}

If we remove the time constant difference excess, we find in the  interval 
minutes 4.5 through 9.5 of series-B a superposed ratio excess that varies smoothly 
between 31 and 38\,Mm with a flat maximum of 
$V_{macro}($Fe\,\textsc{ii}$)/V_{macro}($He\,\textsc{i}$)\approx1.25$ (yellow line in
Fig.\,11). In series-A we find a ratio excess with a maximum of 1.35 (full green line
in Fig.\,11). Both excess values are close to that of 1.22 in Wiehr et al. (2019).

In those data the ratio excess was observed along the complete slit length. 
The present data, however, show a spatial restriction of the ratio excess to 3 - 8\,Mm
height. Since our observations were performed at a single cut through the prominence
from a fixed slit position, we have no information about the two-dimensional spatial
distribution of the ion velocity excess. In both studies we do not find phase shifts
between the Fe$\,\textsc{ii}$ and He$\,\textsc{i}$ velocities, in agreement with Khomenko
et al. (2016) and Wiehr et al. (2019).

The pronounced high frequency oscillations with 22\,s period show an ion velocity excess
with a maximum of 1.7 in a spatial interval of about 3\,Mm. These short periods occur
only in the time interval of minute 3 through 6 in series-A (cf. Fig.\,9). They do not
vary with height, indicating stationary waves. Short period oscillations are rarely
observed in solar prominences. Balthasar et al. (1993) find 30\,sec periods in a quiescent
prominence, and Locans et al. (1983) observe 43\,sec periods in a coronal arch system
around a prominence. In contrast to these high-frequency oscillations with several
pronounced periods of decreasing amplitude (i.e. damped cf. Fig.\,9), do the quasi-periodic
velocity perturbations of longer time scale (cf. Figs.\,4 and 8) show only few extrema
with nearly amplitude. This quasi-periodic character is typical for most velocity time
series observations in quiescent prominences (cf. Wiehr, 2004).

The observation of a spatial and temporal restriction of the ion excess indicates an
interaction of a structured velocity field with the prominence magnetic field. An
impressive view of such complex velocity field in a prominence is seen in the Hinode
observations by Berger et al. (2010). The ion velocity excess will occur only under
particular conditions of such a chaotic velocity field, in accordance with the findings of
Khomenko et al. (2016) of an excess only in short-lived transients with high velocities.
These authors argue that the balance between ions and neutrals "is usually lost a
locations with large individual velocities or large spatial or temporal gradients", and
further that they "are smoothed by interactions on relatively short timescales (typically
of the order of minutes)". The here described observations of the ion-neutral velocity
ratio excesses are indeed limited in space and time.

The present observations were made $\approx$\,100\,min before the sudden disappearance 
of the prominence, which may announce itself by the increasingly chaotic velocity field.
In contrast, the finding of a systematic excess during the full observing time through
the whole prominence  time (Wiehr et al., 2019) was obtained from a long-living quiescent
prominence. Hence the dynamic behavior of the prominence seems to play an essential role
for the detectability of an ion velocity excess over neutrals.

\eject
\section{References}

Ballester,\,J.\,L.; Alexejev,\,I.; Collados,\,M.; Downes,\,T.; Pfaff,\,R.\,F.; 
Gilbert,\,H.; Khodachenko,\,M.; Khomenko,\,E.; Shaikhislamov,\,I.\,F.; 
Soler,\,R.; V\'azquez-Semadeni,\,E.; Zaqarashvili,\,T.: 2018, SSRv\,214, 58
(DOI: 10.1007/s11214-018-0485-6)
\ \\

Balthasar, H.: 2003, Sol. Phys. 218, 85 (DOI: 10.1023/B:SOLA.0000013028.11720.0d) 
\ \\

Balthasar, H., Wiehr, E., Schleicher, H., W\"ohl, H.: 1993, A\&A 277, 635
\ \\

Berger, T. E., Slater, G., Hurlburt, N. Shine, R., Tarbel, T., Title, A., Lites, B. W.,
Okamoto, T. J., Ichimoto, K., Katsukawa, Y., Magara, T., Suematsu, Y., Shimizu, T.:
2010, ApJ 716, 1288 (DOI: 10.1088/0004-637X/716/2/1288)
\ \\

Gilbert, H. R., Hansteen, V. H., Holzer, T. E.: 2002, ApJ 577, 464 (DOI: 10.1086/342165)
\ \\

Khomenko, E., Collados, M., Diaz, A.J.: 2016, ApJ 823, 132 (DOI: 10.3847/0004-637X/823/2/132)
\ \\

Locans, V., Zhugzda, D., Koutchmy, S.: 1983, A\&A 120, 185
\ \\

Ramelli, R., Stellmacher, G., Wiehr, E., Bianda, M.:  2012, Sol. Phys. 281, 697
(DOI: 10.1007/s11207-012-0118-2 )
\ \\

Stellmacher, G., Wiehr, E.: 2015, A\&A 581, 141 (DOI: 10.1051/0004-6361/201322781)
\ \\

Wiehr,E.: 2004, Proceedings of 'SOHO 13 - Waves, Oscillations and Small-Scale
Transient Events in the Solar Atmosphere: A Joint View from SOHO and TRACE'.
29 September - 3 October 2003, Palma de Mallorca, Balearic Islands, Spain
(ESA SP-547, January 2004). 
\ \\

Wiehr, E., Stellmacher, G., Bianda, M.: 2019, ApJ 873, 125 (DOI: 10.3847/1538-4357/ab04a4)

\end{document}